\documentclass[conference]{IEEEtran}
\IEEEoverridecommandlockouts
\usepackage{cite}
\usepackage{amsmath,amssymb,amsfonts}
\usepackage{algorithmic}
\usepackage{graphicx}
\usepackage{textcomp}
\usepackage{xcolor}
\usepackage{epstopdf}
\usepackage{subfigure}

\def\BibTeX{{\rm B\kern-.05em{\sc i\kern-.025em b}\kern-.08em
    T\kern-.1667em\lower.7ex\hbox{E}\kern-.125emX}}
\begin{document}

\title{An Improved Ray Tracing Acceleration Algorithm Based on Bounding Volume Hierarchies\\
	
{\footnotesize }
}

%\author{\IEEEauthorblockN{Chen Wang}
%\IEEEauthorblockA{\textit{School of Information Science and Engineering} \\
%\textit{Southeast University}\\
%Nanjing, China \\
%1437229835@qq.com}
%\and
%\and
%\and
%\and
%\IEEEauthorblockA{\textit{dept. name of organization (of Aff.)} \\%
%\and
\author{Chen Wang\textsuperscript{1}, Yinghua Wang\textsuperscript{2}, Yuxiao Li\textsuperscript{1,2}, Jialing Huang\textsuperscript{1,2}, Jie Huang\textsuperscript{1,2}, and Cheng-Xiang Wang\textsuperscript{1,2*}\\
	\textsuperscript{1}{National Mobile Communications Research Laboratory,}\\
	{School of Information Science and Engineering, Southeast University, Nanjing 210096, China.}\\
	\textsuperscript{2}{Purple Mountain Laboratories, Nanjing 211111, China.}\\
	\textsuperscript{*}{Corresponding Author: Cheng-Xiang Wang}\\
	%	$^3$\small{Shandong Provincial Key Lab of Wireless Communication Technologies,}\\
	%	\small{School of Information Science and Engineering, Shandong University, Qingdao, Shandong, 266237, China.} \\
	Email: wangc67@foxmail.com, wangyinghua@pmlabs.com.cn, \{yuxli, jlhuang, j\_huang, chxwang\}@seu.edu.cn\\
}

\maketitle

\begin{abstract}
Ray tracing is an efficient channel modeling method. However, the traditional ray tracing method has high computation complexity. To solve this problem, an improved bounding volume hierarchies (BVH) algorithm is proposed in this paper. Based on surface area heuristic (SAH) and spatial distance, the proposed algorithm can effectively reduce the number of unnecessary intersection tests between ray and triangular facets. In addition, the algorithm fully considers the influence of ray action range, which can not only make up for the defects of spatial division based on uniform grid method and k-dimensional (KD) tree, but also solve the problem of unsatisfactory spatial division based on traditional BVH algorithm. The simulation results show that compared with the traditional BVH algorithm, the proposed algorithm can improve the computation efficiency by 20\% to~35\% while ensuring the computation accuracy. 
\end{abstract}


\section{Introduction}
Based on geometrical optics and uniform theory of diffraction, ray tracing method can provide accurate prediction results. In recent years, the method has been widely used in the prediction of radio wave propagation characteristics in microcellular environments \cite{b1, b2, b3}. 
However, due to its high computation complexity, even for the simple scenario, ray tracing still faces the challenge of low computation efficiency~\cite{b4}. Therefore, how to use ray tracing method to accurately and efficiently predict propagation characteristics of radio waves becomes particularly important.

In the traditional ray tracing method, each ray needs to be judged with all building primitives in the scenario to determine whether the ray is blocked by obstacles. In fact, in a complex scenario, most of the primitives are far away from the ray that only a few may intersect the ray, so it is not necessary to traverse all of them. Spatial division algorithm has been used to select obstacles within a certain range for testing, instead of testing all obstacles \cite{b5}. In this way, irrelevant primitives can be quickly removed, thus collisions between ray and buildings can be quickly and accurately determined by traversing only a small number of primitives.

There are many common spatial division algorithms. Uniform grid divides the whole space into $n$ subspaces of the same size \cite{b6, b7}. Octree recursively divides the space into eight parts using three mutually perpendicular planes \cite{b8}. Binary space partitioning (BSP) recursively divides the space into two subspaces by planes with arbitrary orientations and arbitrary positions \cite{b9}.~KD tree is similar to BSP except that an axis-aligned split plane is selected each time \cite{b10,b11,b12}. They are all acceleration structures based on space division, i.e., the whole scenario is divided into several subspaces. Therefore, the same primitives may be divided into different subspaces, which will lead to multiple references to the same primitives contained in different subspaces during intersection testing, reducing the computation efficiency. 

BVH algorithm is based on object division, which can make up the defects of the above mentioned algorithms \cite{b13}. However, the traditional BVH algorithm is inefficient when the primitives are distributed unevenly. To solve this problem, Salmon and Goldsmith proposed the widely used SAH algorithm and analyzed the optimization of BVH construction based on SAH heuristic strategy \cite{b14}. In \cite{b15}, the authors proposed a split clipping algorithm for BVH. The algorithm relaxed the restriction that only one primitive can be stored in each BVH leaf node. 
The edge volume heuristic algorithm was proposed to reduce the overlap between nodes \cite{b16}. A spatial splits volume hierarchy was proposed in \cite{b17}. First, the SAH algorithm was used to find the most suitable candidate positions for primitive division, and then the candidate positions for spatial division were found through chopped binning. However, these algorithms do not fully consider the impact of the distance between the ray and the bounding volume on the scenario~segmentation. In fact, due to the limitation of radiation range, the possibility of ray intersecting with distant buildings is very low, thus the spatial distance between ray and buildings is of great importance for spatial segmentation. 

In this paper, an improved BVH algorithm is proposed, which combines SAH and spatial distance. The algorithm can be used to solve the problem of invalid overlap of bounding volumes caused by imperfect division. In addition, the proposed algorithm is applied to the intersection test between rays and buildings, which further improves the computation efficiency of ray tracing.

The remainder of this paper is organized as follows. Section~\uppercase\expandafter{\romannumeral2} introduces the improved BVH algorithm based on~SAH~and spatial distance. Section \uppercase\expandafter{\romannumeral3} introduces the whole process of ray tracing method. Simulation results and analysis are presented in Section \uppercase\expandafter{\romannumeral4}. Conclusions are drawn in Section~\uppercase\expandafter{\romannumeral5}.

\section{The Improved BVH Algorithm Based on SAH and Spatial Distance}
\subsection{The Traditional BVH Algorithm}
In order to describe the information of irregular objects, complex three-dimensional (3D) ray tracing simulation scenarios are usually stored as multiple triangular facets. Therefore, the intersection test between rays and objects can be transformed into the intersection test between rays and triangular~facets.

The basic idea of BVH algorithm to improve the computation efficiency of ray tracing is to surround a certain number of triangular facets with a slightly larger and simpler geometry. First, judge whether the ray intersects with the bounding volume. If the ray intersects with the bounding volume, it is necessary to further calculate whether the ray intersects the triangular facets in the bounding volume. If not, there is no need to test the intersection between the ray and the triangular facets in the bounding volume.
\subsection{The Improved BVH Algorithm}
BVH algorithm needs to select the appropriate division strategy to split space. The division strategy based on SAH algorithm evaluates the intersection cost and traversal cost of each division \cite{b18}. The purpose is to find the division with the lowest cost, i.e., the most reasonable division under the current situation. When a ray intersects with the scenario, the ray may hit either the left node bounding volume $A$ or the right node bounding volume $B$. It can be estimated by probability to determine which area the ray will hit. The main idea of SAH algorithm is that the larger the surface area of the bounding volume, the more likely it is to collide with rays. Therefore, the probability of ray hitting the bounding volume can be estimated by the surface area of the bounding volume,~i.e.,
\begin{equation}
	{c}(A,B)=\frac{{S}(A)}{{S}(C)}{k}_{A}{t}_{i}+\frac{{S}(B)}{{S}(C)}{k}_{B}{t}_{i}+{t}_{trav}
\end{equation}
where $S(A)$ and $S(B)$ represent the surface areas of the left and right node bounding volumes $A$ and $B$, respectively. $S(C)$ represents the surface area of the parent node bounding volume $C$, ${t}_{trav}$ represents the BVH tree structure construction cost, ${t}_{i}$ represents the intersection cost, ${k}_{A}$ and ${k}_{B}$ represent the number of triangular facets in bounding volumes $A$ and $B$,~respectively.

The SAH algorithm estimates whether the ray can intersect a bounding volume according to the surface area of it. However, considering that the range of a ray is limited, whether the ray can intersect with the bounding volume depends not only on the surface area of the bounding volume, but also on the distance from the ray to it. Therefore, based on (1) and considering the influence of spatial distance, this paper proposes a hybrid cost function of SAH and spatial distance, which can be expressed as
\begin{equation}
	{c}(A,B)={\alpha}(\frac{{S}(A)}{{S}(C)}{k}_{A}{t}_{i}+\frac{{S}(B)}{{S}(C)}{k}_{B}{t}_{i})+(1-\alpha){d}^2+{t}_{trav}
\end{equation}
where $d$ is the distance between the center of gravity of the bounding volume and the ray source point, ${\alpha}$ is the weight coefficient between the surface area of the bounding volume and the spatial distance, and the value range is [0,1]. When $\alpha$ is 1, only the surface area of the bounding volume is considered. In the case that $\alpha$ is 0, only the spatial distance is considered. The value of $\alpha$ can be determined according to the specific~scenarios.

In this way, the improved BVH division function not only takes into account the surface area and the overlap degree of subnode bounding volumes, but also the spatial distance. It can overcome the situation that the number of objects in the left subtree and the right subtree is too different and the bounding volumes overlap excessively.

\subsection{Structure of the BVH Tree }
The intersection test of ray and bounding volume needs to be fast, and the bounding volume needs to be very compact. Axis aligned bounding box is actually a simple cuboid, and each of its faces is parallel to the plane of the coordinate axis, thus the intersection of the ray and the bounding volume is easy to calculate \cite{b19}. Therefore, this paper considers using axis aligned bounding box to construct BVH tree.

When determining the bounding volume, it is only necessary to record the minimum and maximum values of the bounding volume on the plane of each coordinate axis. All points in the bounding volume must meet the following conditions
\begin{equation}\label{eq: no1}
	\begin{aligned}
		&{{x}_{min}}<{{x}}<{{x}_{max}} \\ 	
		&{{y}_{min}}<{{y}}<{{y}_{max}} \\ 
		&{{z}_{min}}<{{z}}<{{z}_{max}}
	\end{aligned}
\end{equation}
where (${x}_{min}$, ${y}_{min}$, ${z}_{min}$) represents the minimum coordinate value of the bounding volume on the coordinate axis, (${x}_{max}$,~${y}_{max}$, ${z}_{max}$) represents the maximum coordinate value.

When building a BVH tree, if the number of triangular facets in a bounding volume is too small, the number of tree layers will increase, the tree building time will be long, and the bounding volume may be highly overlapped. If there are too many triangular facets in a bounding volume, the tree building time will reduce, but the number of intersection tests between rays and triangular facets will be greatly increased, and the ray tracing time will be longer. Therefore, before building the BVH tree, it is necessary to reasonably set the number of triangular facets in the bounding volume, so as to effectively improve the efficiency of tree building. The structure of BVH tree is shown in Fig.~1.

\begin{figure}[ht]
	\centerline{\includegraphics[width=0.42\textwidth]{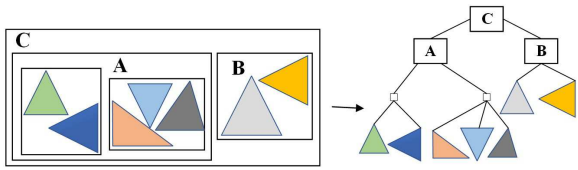}}
	\caption{Bounding volume hierarchies structure.}
	\label{fig}
\end{figure}
The BVH tree is recursively constructed from top to bottom, as shown in Fig. 2. The main flow of the algorithm is as~follows.
\begin{itemize}
	\item Import the triangular facets and their corresponding indices in the scenario construction.
	\item Calculate the bounding volume that can contain all the triangular facets in the scenario.
	\item Select the optimal division strategy based on the improved BVH algorithm, and divide all triangular facets into two parts (i.e., left and right child nodes).
	\item Recursively, repeat the previous for the part of the child nodes, until the number of triangular facets in the current bounding volume is less than the preset maximum number of triangular facets that can be stored in a node. Then the last triangular facets information is stored on the leaf node to end the construction of the BVH tree.
\end{itemize}

It can be seen that if the number of triangular facets in the bounding volume is greater than the threshold, the tree will be recursively constructed until all boundary volumes meet the conditions for establishing leaf nodes.

\begin{figure}[bp]
	\centerline{\includegraphics[width=0.42\textwidth]{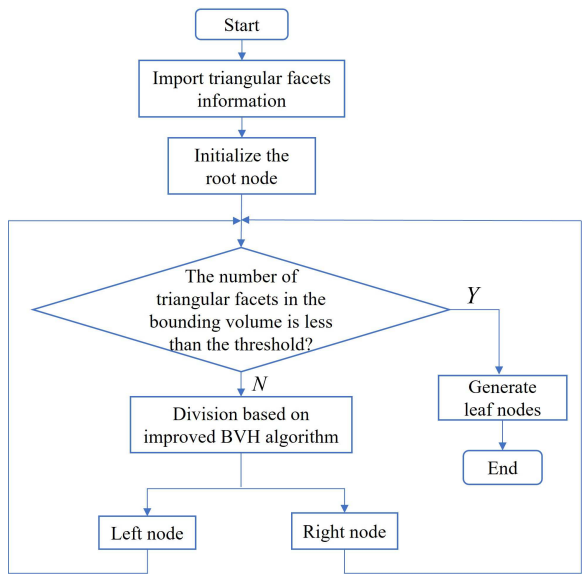}}
	\caption{The flowchart of constructing the BVH tree.}
	\label{fig}
\end{figure}
\section{Ray Tracing Method Based on the Improved BVH Algorithm}
\subsection{Ray Launching}\label{AA}
All rays emitted from the source point must be divided as evenly as possible. According to the shooting and bouncing ray (SBR) algorithm \cite{b20}, this paper considers dividing the sphere into a regular icosahedron composed of 20 identical equilateral triangular faces and 12 vertices, as shown in Fig.~3.

\begin{figure}[tp]
	\centerline{\includegraphics[width=0.3\textwidth]{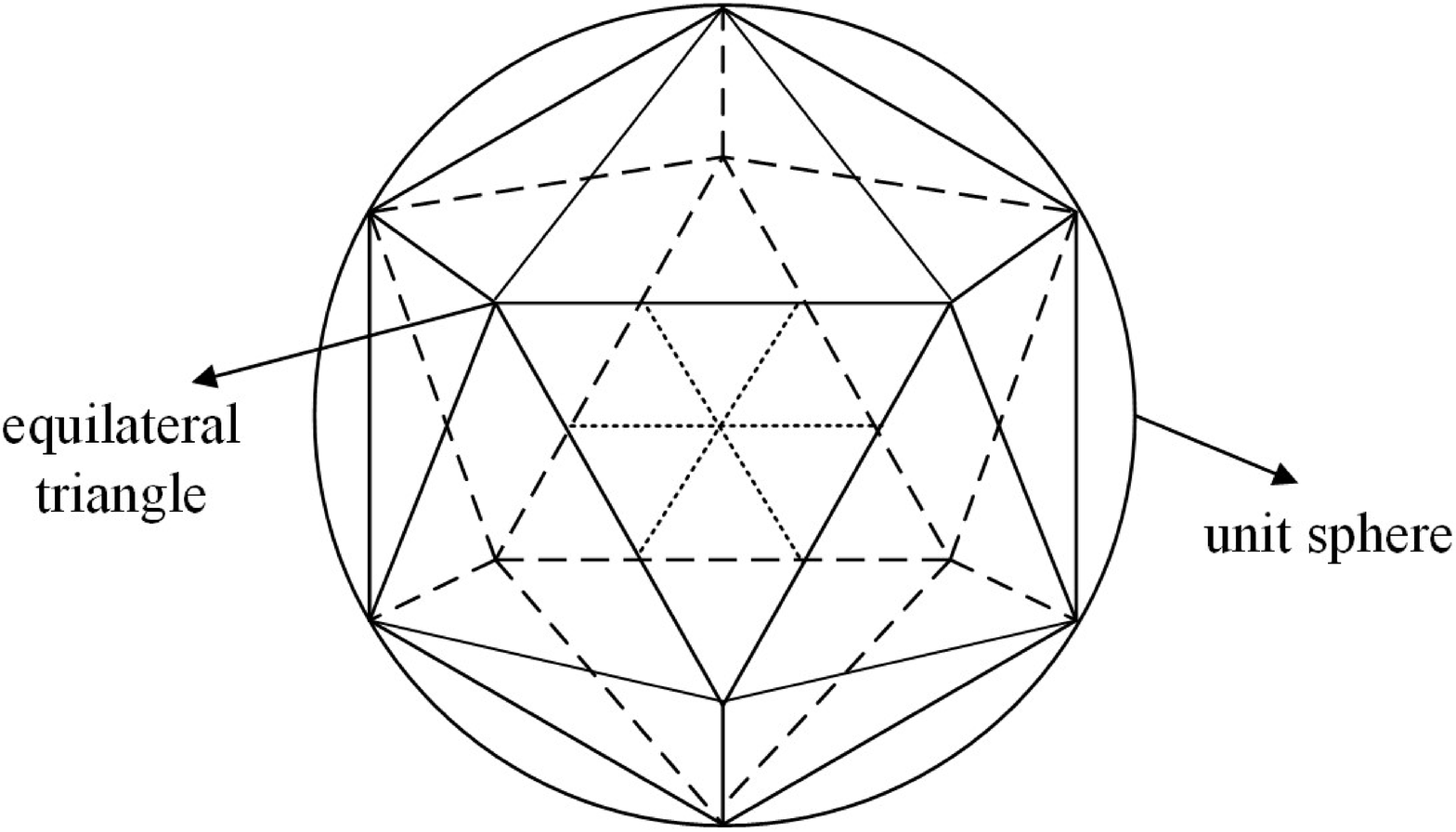}}
	\caption{Regular icosahedron.}
	\label{fig}
\end{figure}
To obtain more and densier ray beams, each equilateral triangle can be further subdivided to form more and smaller equilateral triangles. Finally, all the initial ray beams can be obtained by connecting the source point to the vertices of these~triangles.
\subsection{Ray Tracing}
When the ray is emitted from the source point, it may reach the field point directly, or it may arrive at the field point after several processes of reflection and diffraction. The intersection test is the key process of ray tracing, and the algorithm flow is shown in Fig. 4. Because the buildings are stored in the format of triangular facets, and the triangular facets are enclosed in the BVH tree, the intersection test of the ray and the building can be transformed into the collision detection of the ray and the BVH tree.  

\begin{figure}[bp]
	\centerline{\includegraphics[width=0.31\textwidth]{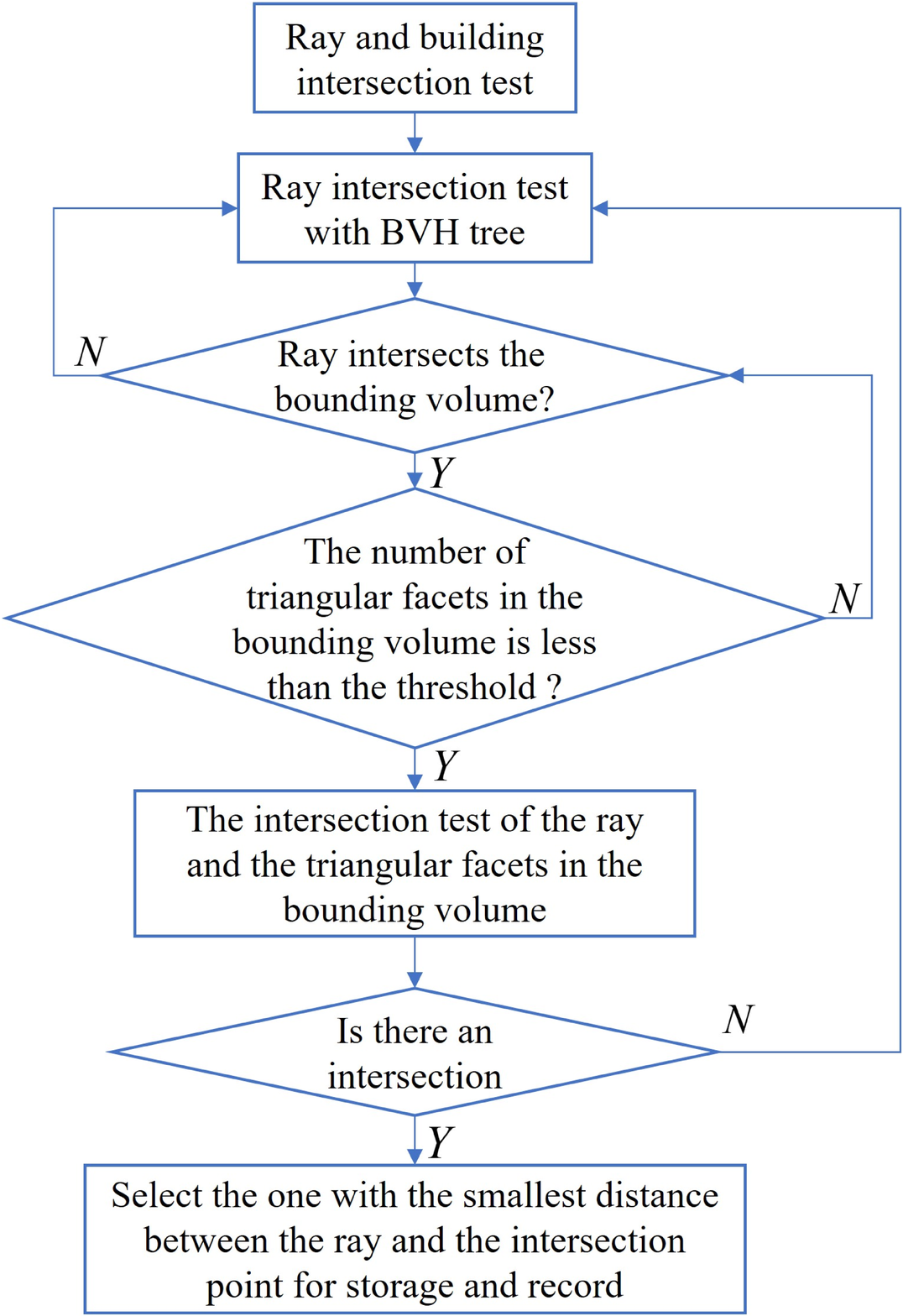}}
	\caption{The flowchart of intersection test.}
	\label{fig}
\end{figure}
The process of collision detection between ray and the BVH tree is summarized as follows. 
\begin{itemize}
	\item Find the bounding volume of triangular facets.
	\item Judge whether ray intersects with the bounding volume.
	\item If they intersect, the bounding volume intersecting the ray is further tracked to determine which sub-bounding volume in the bounding volume will intersect with the~ray.
	\item Repeat the previous process until the number of triangular facets in the bounding volume is less than the threshold,~i.e., traverse the leaf nodes on the binary tree.
	\item Calculate the intersection point between the ray and the triangular facets contained in the leaf node, select the one with the smallest distance between the ray and the intersection point, store the corresponding results and~return.
\end{itemize}

When ray collides with triangular facets, it will generate reflected and diffracted rays. In the process of ray tracing, the propagation path of each ray is obtained by tracing the rays in different emission directions. When the energy of the signal decays to a certain threshold or the simulation order reaches the set order of reflection and diffraction, the ray tracing process will cease.
\subsection{Ray Reception}
After all the rays are traced, the SBR algorithm uses the concept of the receiving sphere to determine whether the rays can be received by the field point, then coherently superimposes the received rays to obtain the total energy at the field point. The radius $r$ of the receiving sphere under ray tracing is generally expressed as
\begin{equation}
r\thickapprox\frac{l\delta}{\sqrt{3}}
\end{equation}
where $l$ is the actual path length of the ray between the source point and the field point, ${\delta}$ is the included angle between two adjacent rays at the source point.
\section{Simulation Results and Analysis}
\subsection{Indoor and Outdoor Simulation Parameters Setting}
We built indoor and outdoor simulation scenarios through SketchUp software. The dimension of the simulated indoor study-room scenario is {8}$\times${6}$\times${3.5}~{m}$^{3}$, as shown in Fig.~5. The simulation scenario is furnished with twelve desks, six windows, one door, and other common office furniture. Door and desks are made of wood, walls are made of concrete, and window guardrails are made of metal. 

This scenario contains 2778 triangular facets. We use ($x$,~$y$,~$z$) to represent position coordinates in meter. The transmitter (Tx) is located at (1.46, 2.42, 2.1), while the receiver (Rx) is located at (1.2, 1.2, 1.5). The simulation frequency is set at 2.4 GHz, and the number of triangular facets in the bounding volume is set to 250. For the indoor scenario, the distance between the bounding volume and Tx is very short, so the surface area of the bounding volume has large impact on the overlap of the bounding volumes. In this case, we set the coefficient $\alpha$ in (2) to 0.7.

\begin{figure}[tp]
	\centerline{\includegraphics[width=0.24\textwidth]{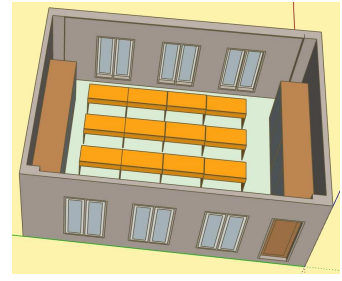}}
	\caption{Indoor study-room scenario.}
	\label{fig}
\end{figure}
Fig. 6 shows the simulated outdoor scenario of Nanjing University of Posts and Telecommunications, which contains 7608 triangular facets. The simulation scenario includes canteens, dormitories, research activities buildings, teaching buildings, libraries, playgrounds, and other buildings. Tx is located at (450, 450, 10), while Rx is located at (450, 550,~25). The simulation frequency is set at 2.4 GHz. The number of triangular facets in the bounding volume is set to 820, and $\alpha$ in (2) is set to 0.4.
\begin{figure}[tp]
	\centerline{\includegraphics[width=0.45\textwidth]{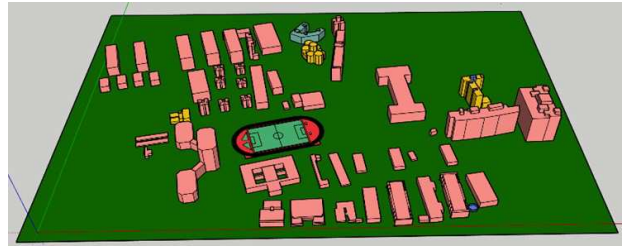}}
	\caption{Outdoor campus scenario.}
	\label{fig}
\end{figure}
\subsection{Analysis of Indoor and Outdoor Simulation Results}
In the simulated indoor scenario, when the threshold of the triangular facets in the bounding volume is set to 250, we compare the scenario division results under different BVH algorithms. The result of scenario division by using the traditional BVH algorithm is shown in Fig. 7(a), and that by using the proposed BVH algorithm is shown in Fig. 7(b). Through comparison, we can find that the algorithm proposed in this paper is more effective for space division, and the bounding volume can effectively surround desks, cabinets and other furniture. At the same time, we also compare the spatial division results of the simulated outdoor scenario when the threshold is 820, as shown in Fig. 8. It can be seen that by using the proposed algorithm, the bounding volume can surround buildings with different shapes in a regular and orderly way. In summary, the proposed algorithm can solve the problem of invalid overlap of bounding volumes caused by the uneven distribution of triangular facets in the traditional BVH algorithm.

\begin{figure}[bp]
	\centering
	\subfigure[Traditional BVH algorithm.]{
			\includegraphics[width=0.2\textwidth]{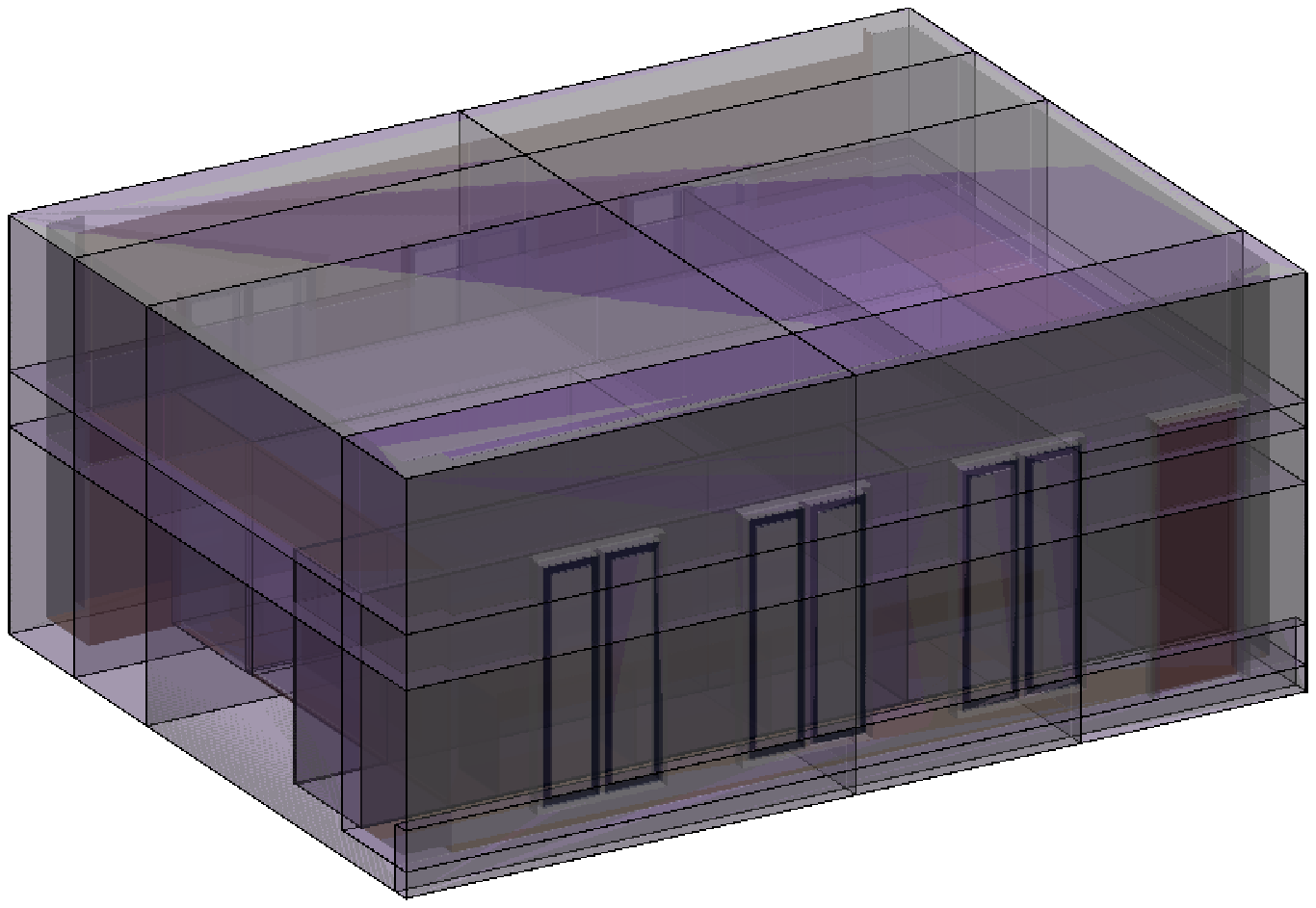}
			%\caption{fig1}
		}
		\quad
		\subfigure[Proposed BVH algorithm.]{
			\includegraphics[width=0.2\textwidth]{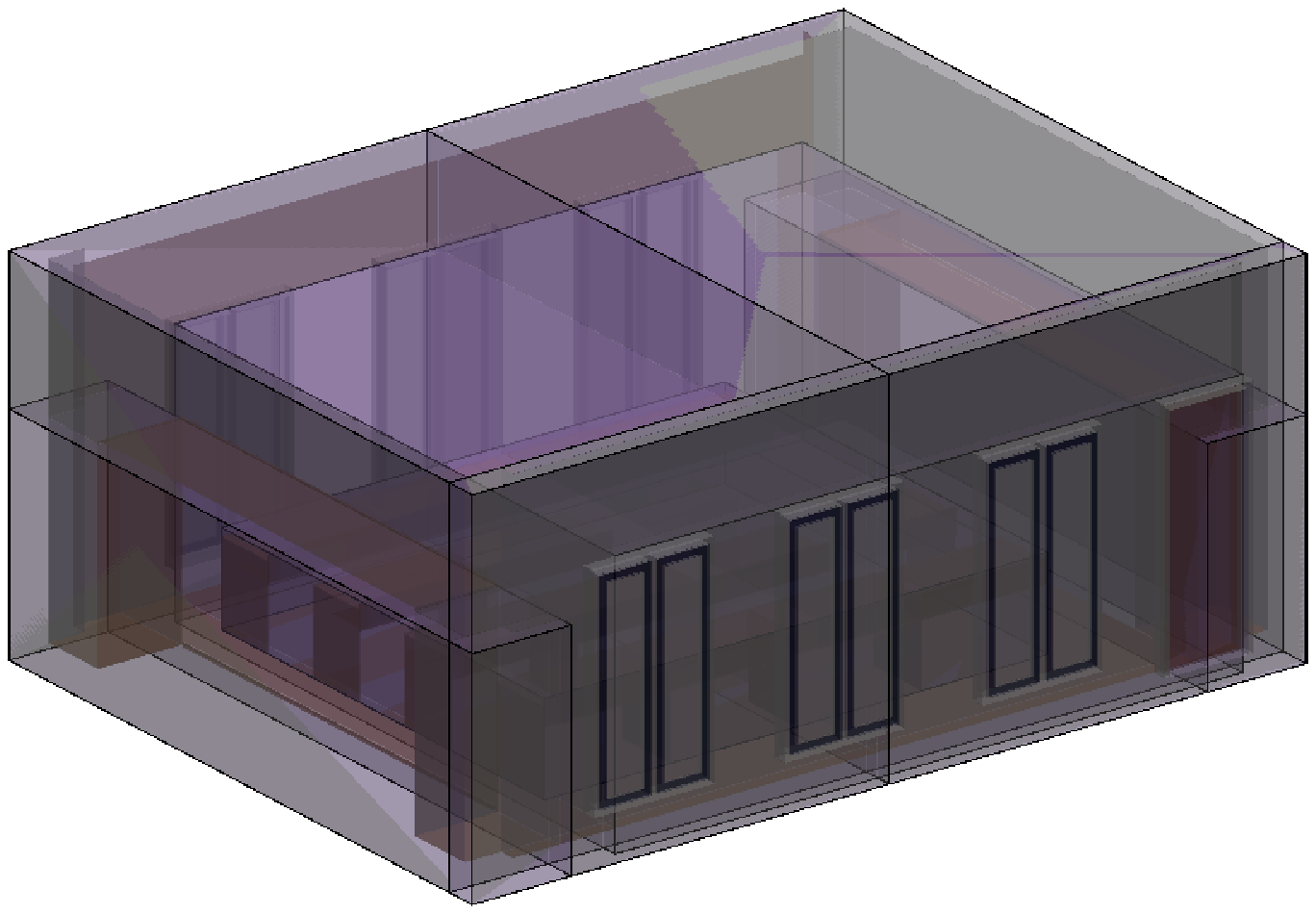}
		}
		\caption{Indoor scenario division results under different BVH algorithms.}
\end{figure}
\begin{figure}[tp]
	\centering
	\subfigure[Traditional BVH algorithm.]{
		\includegraphics[width=0.21\textwidth]{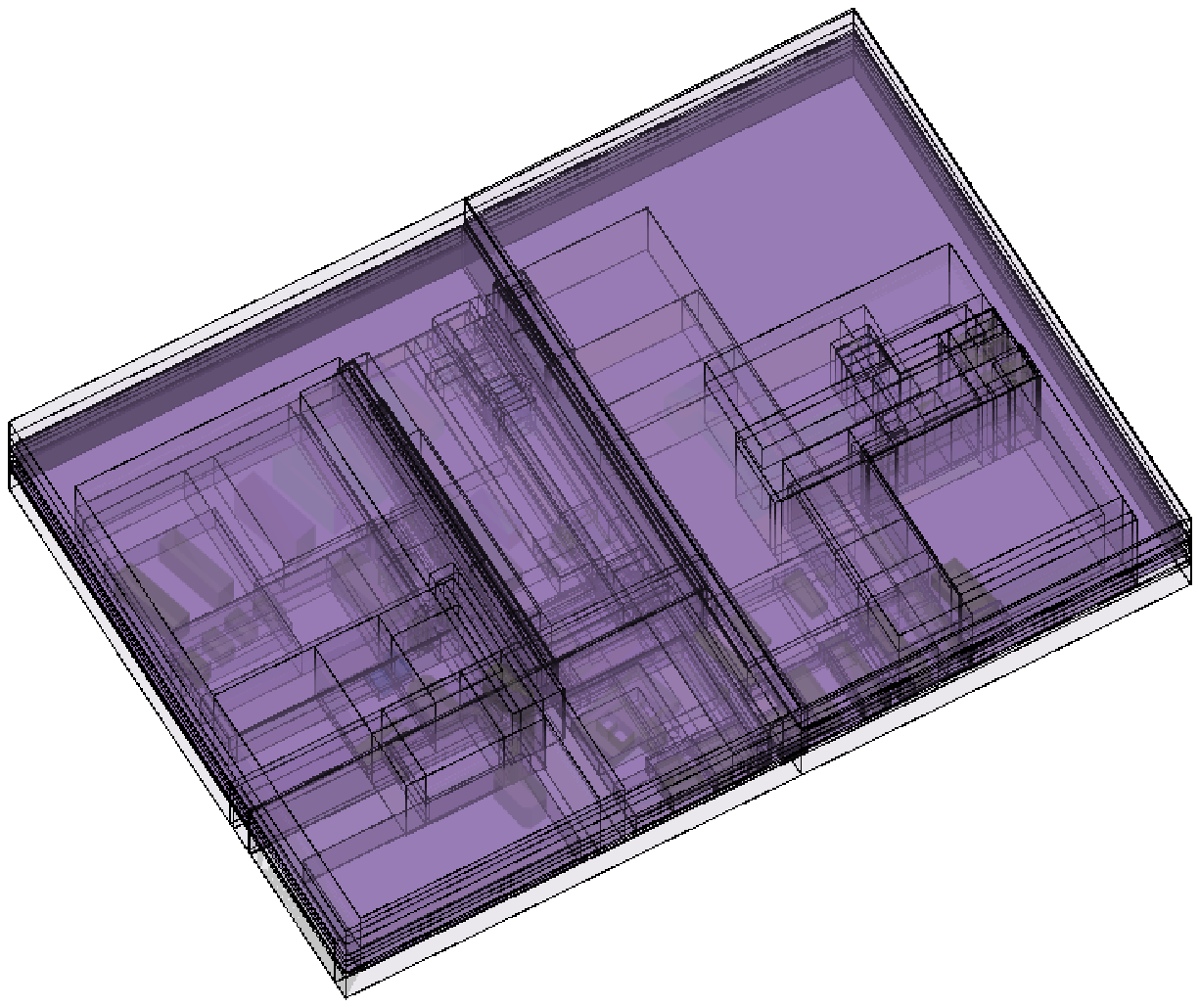}
		%\caption{fig1}
	}
	\quad
	\subfigure[Proposed BVH algorithm.]{
		\includegraphics[width=0.21\textwidth]{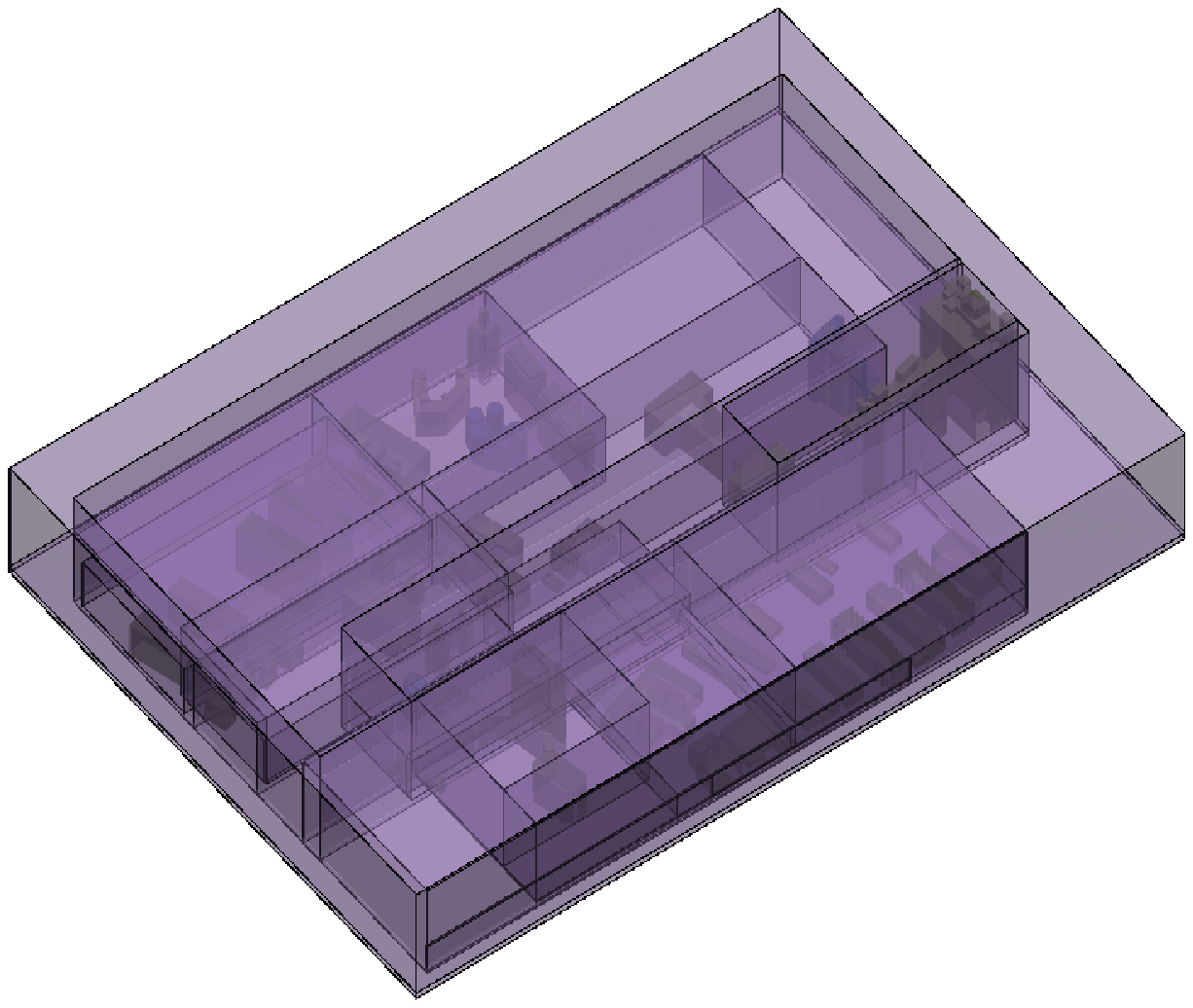}
	}
	\caption{Outdoor scenario division results under different BVH algorithms.}
\end{figure}

\begin{figure}[tp]
	\centering
	\subfigure[Indoor scenario simulation.]{
		\includegraphics[width=0.45\textwidth]{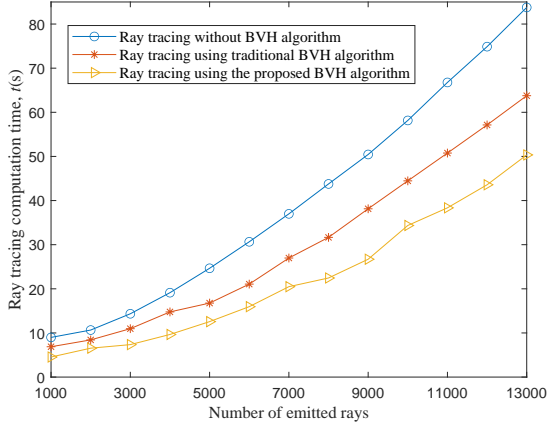}
		%\caption{fig1}
	}
	\quad
	\subfigure[Outdoor scenario simulation.]{
		\includegraphics[width=0.45\textwidth]{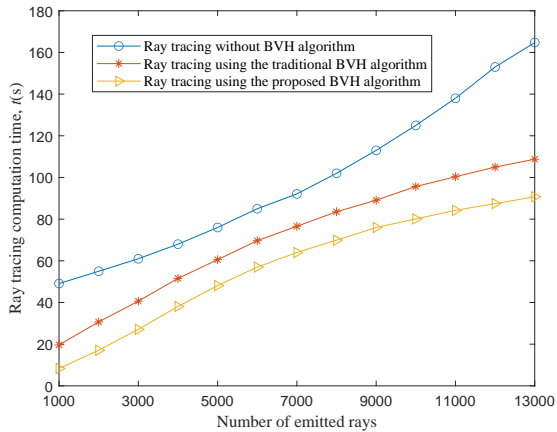}
	}
	\caption{Comparison of different algorithms.}
\end{figure}

Fig. 9 shows the running time of the ray tracing method before and after acceleration under different numbers of emitted rays in indoor and outdoor scenarios. It can be seen that with the increase of the number of emitted rays, the running time also increases correspondingly, and the growth rate gradually accelerates. At the same time, compared with the traditional BVH algorithm, the proposed algorithm can better reduce the computation time. However, different from the simulated indoor study-room scenario, the computation efficiency based on the proposed algorithm is improved more obviously in the simulated outdoor scenario. This is mainly due to the large number of triangular facets in the outdoor campus scenario. The scenario overview is much more complex than the indoor scenario, and the intersection test of rays and triangular facets takes a lot of time. Therefore, the algorithm proposed in this paper can greatly improve the computation efficiency. Meanwhile, the  proposed algorithm is suitable for indoor and outdoor application scenarios.

The reflection order is set to first and second order in the indoor scenario, and third order in the outdoor scenario. The diffraction order of both the indoor and outdoor simulation scenarios is set to 0. The ray paths in indoor and outdoor scenarios before and after using the proposed BVH algorithm are shown in Fig. 10 and Fig.~11. The green line is the direct path from the Tx to the Rx, and the red lines are the reflection ray paths. 
It can be seen that the ray paths obtained after the acceleration are consistent with those before acceleration, indicating that during the ray tracing, only the invalid intersection test process is deleted, and the correct intersection test process is not deleted by mistake. This shows that the proposed algorithm can improve the computation efficiency while ensuring the~accuracy.
\begin{figure}[tp]
	\centering
	\subfigure[First-order reflection without using~the proposed BVH algorithm.]{
		\includegraphics[width=3.9cm]{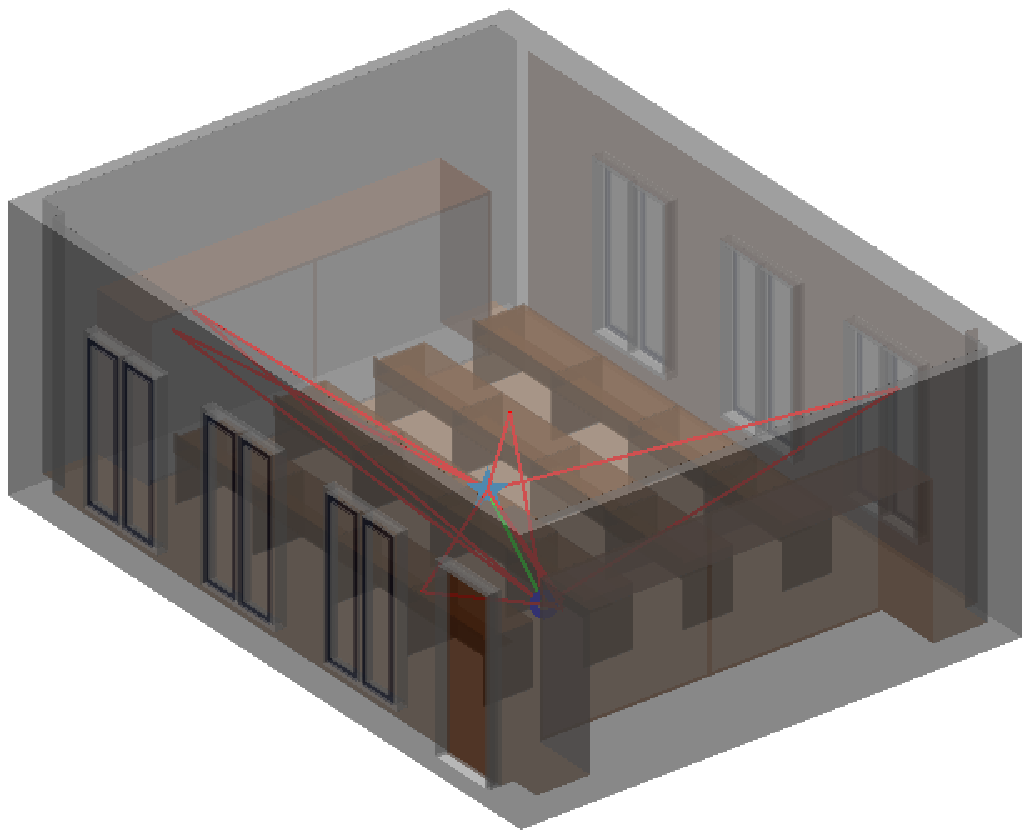}
		%\caption{fig1}
	}
%First-order reflection without using the proposed BVH algorithm
	\quad
	\subfigure[First-order reflection using the proposed BVH algorithm.]{
		\includegraphics[width=3.9cm]{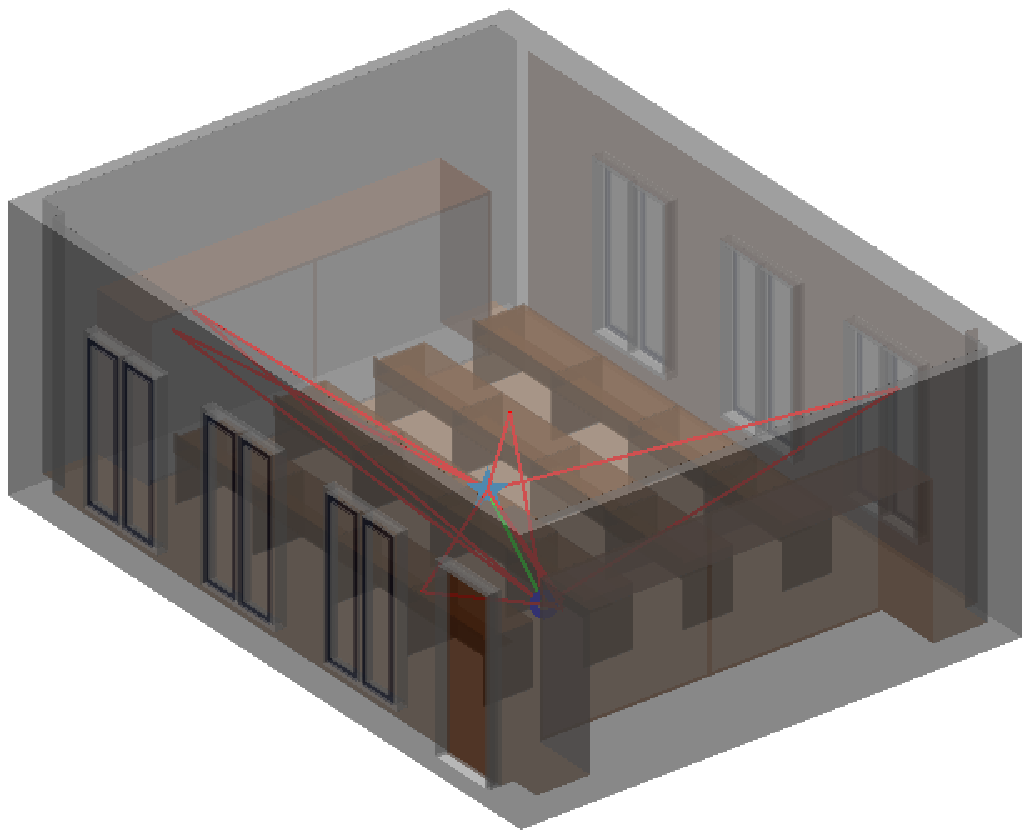}
	}
	\quad
	\subfigure[Second-order reflection without using the proposed BVH algorithm.]{
		\includegraphics[width=3.9cm]{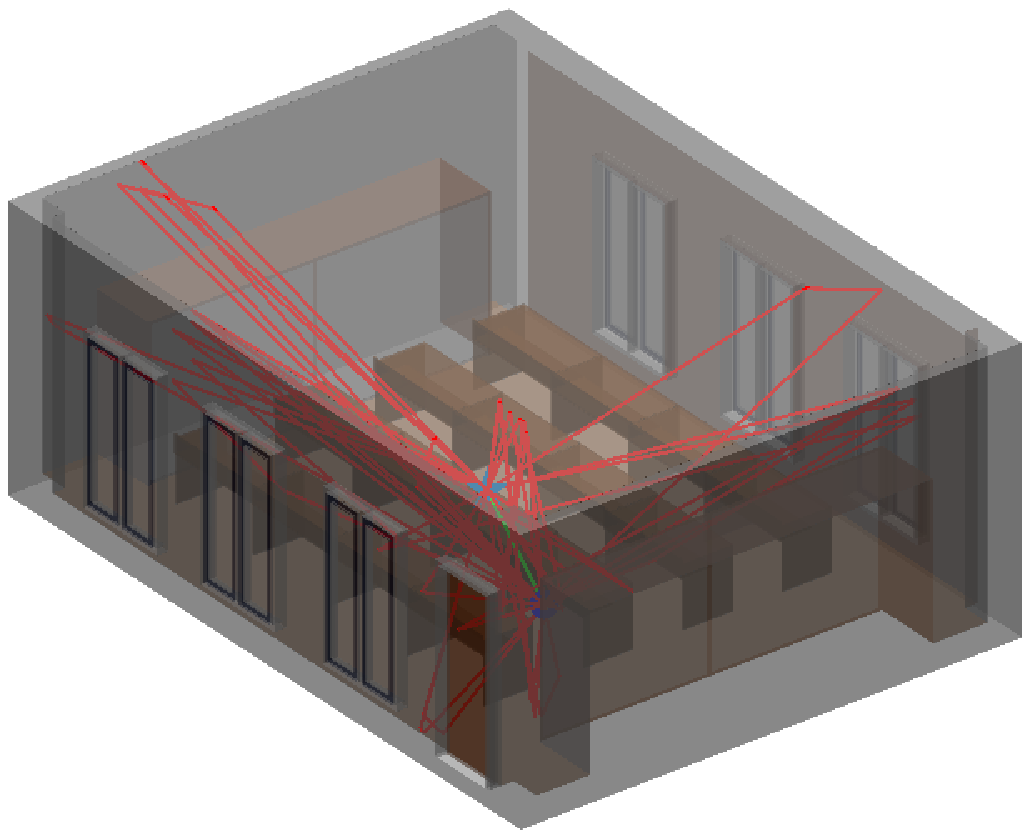}
	}
	\quad
	\subfigure[Second-order reflection using the proposed BVH algorithm.]{
		\includegraphics[width=3.9cm]{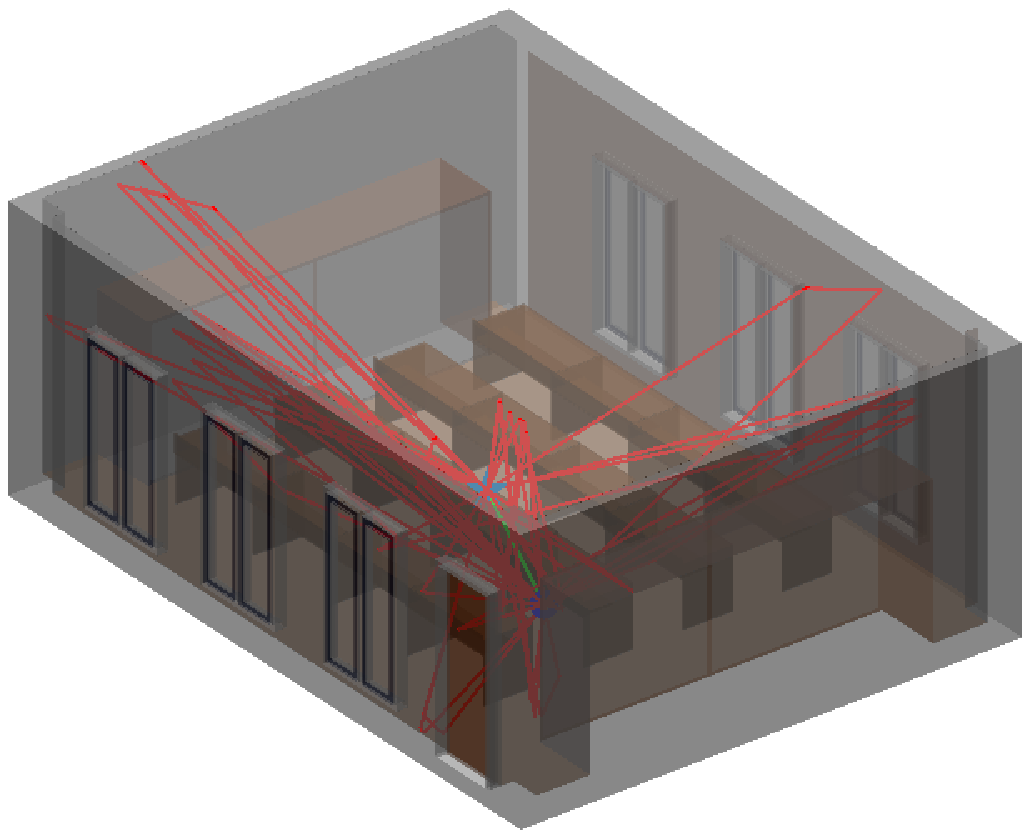}
	}
	\caption{Ray tracing simulation results in indoor scenario.}
\end{figure}
\begin{figure}[th]
	\centering
	\subfigure[Third-order reflection without using the proposed BVH algorithm.]{
		\includegraphics[width=6cm]{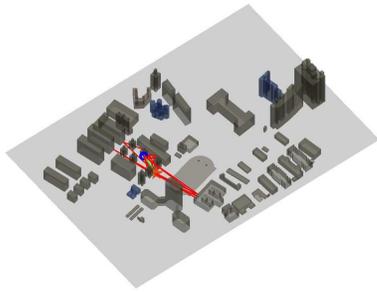}
		%\caption{fig1}
	}
	\quad
	\subfigure[Third-order reflection using the proposed BVH algorithm.]{
		\includegraphics[width=6cm]{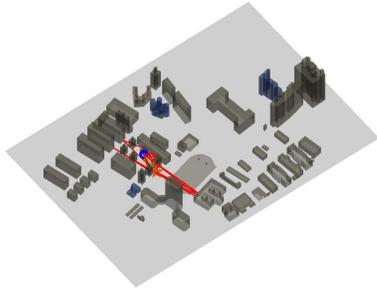}
	}
	\caption{Ray tracing simulation results in outdoor scenario.}
\end{figure}
\section{Conclusions}
In this paper, an improved BVH algorithm has been proposed for ray tracing acceleration. This algorithm can effectively divide space and solve the problem of invalid overlap of bounding volumes in traditional BVH algorithm. At the same time, the algorithm can be used to reduce the useless intersection calculation between ray and triangular facets. The simulation results have indicated that the proposed algorithm can not only improve the computation efficiency, but also ensure the accuracy of ray tracing. In addition, the algorithm can be applied to both indoor and outdoor scenarios. The proposed BVH algorithm is efficient for ray tracing simulation in complex environments, which will be of great importance for the future channel simulation in new~scenarios.

\section*{Acknowledgment}
This work was supported by the National Key R\&D Program of China under Grant 2018YFB1801101, the National Natural Science Foundation of China (NSFC) under Grants 61960206006 and 61901109, the Frontiers Science Center for Mobile Information Communication and Security, the High Level Innovation and Entrepreneurial Research Team Program in Jiangsu, the High Level Innovation and Entrepreneurial Talent Introduction Program in Jiangsu, the Research Fund of National Mobile Communications Research Laboratory, Southeast University, under Grant 2021B02, the EU H2020 RISE TESTBED2 project under Grant 872172, the High Level Innovation and Entrepreneurial Doctor Introduction Program in Jiangsu under Grant JSSCBS20210082, and the Fundamental Research Funds for the Central Universities under Grant 2242022R10067.

\end{document}